\documentclass[draft]{aipproc}
\layoutstyle{6x9}

\begin{document}

\title{Interpreting Negative Probabilities \\in the Context of Double-Slit Interferometry}

\classification{03.65.Ta, 03.70.+k, 14.70.Bh, 42.50.Xa}

\keywords {Negative probability, quantum interference.}

\author{Ghenadie N. Mardari}{
  address={Rutgers University, New Brunswick, NJ 08901, USA}
}

\begin{abstract}
Negative probabilities emerged at intermediate steps in various
attempts to predict the distributions of quantum interference. There
is no consensus on their meaning yet. It has been suggested
(Khrennikov, 1998) that negative probabilities require the existence
of unsuspected correlations between detection events. We evaluate
this claim in light of several representative experiments. In our
assessment, some of its implications are in good agreement with the
data.
\end{abstract}

\maketitle

\section{Introduction}

It is a matter of common sense that an event can either happen or
not. If its occurrence is certain, the rule is that its probability
must be equal to $1$. If it cannot happen, the probability must be
equal to $0$. In all other instances, the probability can have any
value between $0$ and $1$. With this in mind, the idea that an event
can have a negative probability of occurrence is non-sense. Or is
it?

According to Richard Feynman, all negative numbers defy common sense
in real-life situations. An apple trader may begin the day with $5$
apples and end it with $3$, after giving away $10$ and receiving $8$
apples. At no point in time was the trader in possession of a
negative quantity of apples. Still, the analysis of the process as a
whole is greatly simplified by showing that $5-10=-5$ and $-5+8=3$.
Similarly, it must be acceptable to use negative probabilities as
well, so long as they simplify thought and calculations in properly
chosen situations. In his famous paper on negative probability
\cite{feyn}, which he began with the apple trader example, Feynman
says: ``It is not my intention here to contend that the final
probability of a verifiable physical event can be negative. On the
other hand, conditional probabilities and probabilities of imagined
intermediary states may be negative in a calculation of
probabilities of physical events or states'' (p. 238). Indeed,
Feynman went on to conclude that ``all the results of quantum
statistics can be described in classical probability language, with
states replaced by `conditions' defined by a pair of states (or
other variables), provided we accept negative values for these
probabilities'' (p. 248).

Negative probabilities are primarily useful for simplifying
analysis, as confirmed by numerous other developments in modern
physics. Yet, Richard Feynman went beyond mere calculations. He
showed that negative probabilities have meta-theoretical
implications as well. On the one hand, they can tell us something
about the fitness of a theory. If a model predicts negative
probabilities for real detectable states, it must be clearly wrong
or incomplete. On the other hand, they can also tell us something
about Nature. In some contexts, there might be pairs of events, one
of which has an abnormal probability value. This could mean that
only one event is observable during a single act of measurement, in
obvious agreement with Heisenberg's uncertainty principle. The
difficulty, from our point of view, is that interpretation is not
always straightforward. In many cases it is hard to say if a theory
is incomplete, or if Nature simply defies common sense. For example,
how does it help us to know that the distribution of detection
events in a double-slit experiment can be predicted with models that
involve negative probabilities?

Feynman's approach to negative probability was further refined in
the context of double-slit interferometry by Scully, Walther and
Schleich \cite{scully}. They showed that several questions,
formulated by Feynman in abstract terms, can be illustrated
physically with the help of the micromaser which-path detector. This
allowed them to develop a very instructive formalism, and also to
produce new insights into the EPR problem. Of interest to us is the
general interpretation, implicit in this work. It shows that quantum
properties are not reducible to classical phenomena (even though
classical statistics can be used to describe it, as suggested by
Feynman). The non-local interactions of quantum processes must be
fundamental, providing a direct explanation for the fact that
negative probabilities are required. In other words, the latter
cannot be attributed to some sort of undiscovered physical
interaction.

The above comments notwithstanding, Scully, Walther and Schleich did
not attempt to provide any definitive interpretive conclusions. The
problem of explaining the precise meaning of negative probabilities
in the context of double-slit interference remained open. In this
paper we shall focus on the attempt to find a plausible solution in
terms of \textit{p}-adic probability analysis, as formulated by
Khrennikov \cite{andrei4}. This approach is very interesting because
it shows that a classical interaction could still explain quantum
interference. We shall provide a brief overview of the main claims
and the initial difficulties associated with their verification. We
shall also note the limitations of the original proposals and
discuss a new way to verify this model. More importantly, we shall
examine the results of several experiments, which appear to confirm
the main prediction of this approach.

\section{Searching for meaning}

Quantum statistics appeared to be easier to predict than to
interpret. Khrennikov \cite{andrei4} noted that several earlier
models had reproduced the distributions of the double-slit
experiment successfully, but none of them was shown to have a clear
interpretation. In contrast to these approaches, which were
formulated as measure theories, he developed a theory in which
probabilities were defined as relative frequencies. The outcome was
a model with straightforward physical meaning, in part because of
its innovative use of \textit{p}-adic number analysis, which was not
widely used before.

The main finding of this approach is that detection events in a
double-slit experiment obey a logarithmic complexity rule.
Independent random events obey linear complexity rules. Therefore,
individual quanta cannot generate interference independently. They
must display correlated behavior, in order to produce their
well-known distributions. In other words, self-interference cannot
be the reason for the detection pattern. Some sort of interaction,
possibly classical, must occur. This prediction is particularly
remarkable, because it runs against both the Copenhagen
interpretation and the most well-known pilot-wave models. Moreover,
quantum interference was well demonstrated at low rates of emission,
which was widely considered enough to rule out any direct
interaction between single quanta. The concept of self-interference
is pretty much taken for granted today.

Khrennikov showed that direct interaction between the quanta is not
necessary for an interpretation of his results. The correlations
required in this context could be produced by delayed (memory)
effects within measurement systems. Such effects could happen either
at the source of emission, or at the slits, or finally at the
detectors. The proposition was easily testable. Yet, the problem was
that the experiments did not seem to confirm it. Firstly,
interference was demonstrated with pairs of independent
deterministic sources of photons, which emit single pulses on demand
\cite{santori}. This seems to be a convincing argument against
correlation at the source. On the one hand, deterministic sources
can have arbitrary moments of emission, which makes the idea of
correlation between consecutive events implausible. On the other
hand, the same experiment showed that fringe visibility depended
primarily on the parameters of propagation inside the
interferometer. Secondly, there are many types of interferometers
which do not require slits. In some cases, even Young interference
can happen with or without slits (as shown, for example, in ref.
\cite{basano}). It is possible to argue that the slits are
functionally similar to sources in many set-ups, which further
confirms that memory effects at the openings are not essential, even
if conceivably real. Finally, the hypothesis of memory effects at
the detector is undermined by the wide use of modern (other than
screen) detectors. For example, pico-streak cameras satisfy the
requirement of having physical changes in the detector between
detections, and they do not have diminished visibility. In fact,
they can be used to observe important details, such as fringe drift,
extending the class of observable interference phenomena (see for
example, ref. \cite{lour}). More recent detection techniques, such
as fiber-optic scanning, reinforce this conclusion even further,
reducing the likely importance of memory effects during measurement.
There are also experiments with neutron interferometers, quoted by
Khrennikov \textit{et al.} in a follow-up paper \cite{andrei5},
which did not find the expected memory effects. Consequently,
measurement artifacts (such as memory effects) cannot explain
convincingly the peculiarities of non-classical distributions.

Given the above, it seems appropriate to follow the example of
Feynman and ask: what if some of our assumptions are wrong? What if
self-interference is not a real phenomenon, despite its popularity
as an interpretive tool? After all, the simplest way to account for
correlations among detection events is to assume that quanta
interact with each other, even when they are not involved in direct
collisions. Such a hypothesis might seem to go against the grain,
but the relevant question is whether it is possible to formulate a
plausible model for it. As shown elsewhere \cite{gm04,gm05}, such a
model can be developed in a way that is consistent with the relevant
experimental evidence. If quanta are treated as sources of real
waves, they can be shown to produce interference fringes without
colliding directly. In the case of photons, the most important
elements would be the length of pulses and the amplitude of created
waves, which tends to diminish with distance from the source. This
means that the reality of self-interference is testable by checking
for its indicators in critical cases, where the predictions of
different approaches do not converge. According to our analysis, the
expected properties of self-interference did not materialize in
relevant experiments. By implication, Khrennikov's conclusions are
strongly supported by empirical data. This means that quantum
interference could really contain a hidden interaction,
understandable in the language of classical mechanics. By switching
the explanation of negative probabilities from the properties of
individual quanta to the interactions among them, this approach has
reopened the question of completeness of quantum mechanics. At least
in the case of the double-slit experiment, an alternative
interpretation became possible. We review a few of the most relevant
experiments in the following chapter.

\section{Interferometric evidence}

The hypothetical absence of self-interference has several
experimental implications that have already been tested, as part of
unrelated investigations \cite{gm05}. Two of them are especially
relevant for the present discussion. Firstly, correlations must
vanish when their physical preconditions are not met. Below
predictable energy levels, which translate into quantum density per
volume of space-time, interference must become gradually
undetectable. In the case of photons, whose action is proportional
to their duration, these threshold rates must also depend on
pulse-width. Secondly, individual quanta are expected to have
well-defined trajectories. Therefore, interference should persist
even in special settings, in which only one slit is accessible at a
time, provided quanta have alternative access to more than one
opening. At least for the situations that involve photons, both of
these predictions are in agreement with the experimental record.

Several preliminary remarks are in order, before we look at the
data. When a classical wave hits an obstacle with two openings, it
will come out on the other side in the form of two waves, displaying
interference in their area of overlap. In this sense, the original
wave can be described as interfering with itself. Optical
interference confirms the wave properties of light. Given this,
should we expect the quanta of light to interfere with themselves?
If quantized waves were similar to classical waves, then we should
always expect self-interference. Yet, if they were constantly
produced as discrete oscillations by propagating localized sources,
self-interference should be impossible. Every localized source could
only go through one slit, and the waves (defined here as space-time
perturbations) could not be reflected by material obstacles. Both of
these possibilities can be resolved empirically, by looking for
evidence of interference at extremely low energy levels.

It is well-known that a single quantum cannot produce fringes. It
can only produce a detection click. In order to observe
interference, large numbers of coherent photons must be detected.
Still, it is possible to determine if interaction prior to detection
plays any role in the final outcome. For this, we must ensure that
photons do not overlap in transit and see if they display
first-order interference at arbitrary intervals between single
pulses. Firstly, single-photon pulses must have finite duration.
Otherwise, they will always overlap, no matter the time difference
between any two detections. So, they must be chopped, or emitted in
discrete pulses. Secondly, the coherence time of the source must
exceed the interval between any two pulses at emission; otherwise
the main pre-condition for interference will not be met. If these
technical features are guaranteed, and interference persists at any
rate of emission, then we can be confident about the reality of
self-interference.

There are numerous experimental proofs of interference at very low
rates of emission. However, the two conditions mentioned above
(especially pulse-width) were not explicitly enforced in most cases.
And in the few cases, when they appeared to be met, interference
vanished. Usually, insufficient coherence at the source is suspected
in such cases. In our opinion, the evidence does not justify such an
interpretation. For example, Dontsov and Baz \cite{baz} suggested
that discharge tubes, used as sources of photons, cannot produce
coherent light at low levels of excitation. They demonstrated this
by proving that interference vanished, when their source was weak.
Furthermore, interference fringes reappeared, when they increased
the output of the sources by two orders of magnitude. Nevertheless,
when they placed neutral density filters behind the source,
diminishing the rate of detection to the same low levels, visibility
dropped again. It is remarkable that Dontsov and Baz were able to
recover the fringes by moving the filters beyond the interference
volume, in front of the detectors. Thus, interference visibility was
independent from the technical state of the detectors, as well as
from that of the source. The main factor was the number of photons
passing through the interference volume per unit of time.

In a modern demonstration, Ribeiro and collaborators discarded the
idler beam from a source of spontaneous parametric down conversion
(SPDC), and performed a double-slit experiment with the signal beam
\cite{ribeiro}. They used a special set-up to achieve high rates of
emission and controlled the pump (input) beams with neutral density
filters. Narrow-band interference filters were used to screen for
monochromatic detections only. The result was a very clear
demonstration of interference at high emission rates, as well as of
its gradual disappearance at lower rates. Unfortunately, Ribeiro
\textit{et al.} concluded that their source cannot produce coherent
photons at low rates, without testing for alternative explanations.
They could have placed a neutral density filter behind the source,
just like Dontsov and Baz, in order to see if fringes can be
produced by coherent photons at the same low emission rates. So, we
have to look at other experiments for a proper interpretation. In
one experiment, Kim \textit{et al.} controlled the exact interval
between independent signal photons emitted in pairs \cite{kim01}. As
the time-delay between photons was increased, first-order
interference gradually vanished. This shows that the interval
between the quanta was more important than the state of the source
for the final outcome. Though, a possible objection might be that
spontaneous sources cannot ensure phase-coherence, which could be
especially important at large intervals between pulses. Still, there
is another experiment, by Kim and Grice \cite{kim02}, in which
sub-wavelength adjustments of time-delay were achieved. In these
conditions, maximum visibility was made possible by ensuring phase
coherence between interacting photons. Still, interference did not
persist after a threshold interval between photons. This evidence is
sufficient for us to conclude that self-interference did not happen
in a context, in which its preconditions were met. Whatever the
nature of matter waves, they do not seem to produce quantum
interference via self-interaction.

It is also remarkable that interference happened even when quanta
did not overlap in space, provided they were within the threshold
boundaries. This is evidence of remote interaction between them,
consistent with our hypothesis of space-time fluctuations from
localized quantum sources. Furthermore, the interval between
independent detections was larger for Dontsov and Baz \cite{baz}
than for Ribeiro \textit{et al.} \cite{ribeiro}, as expected. Longer
pulses are physically closer to each other at fixed rates of
emission. This element is well supported by the experiment of
Santori \textit{et al.} \cite{santori}. They used quantum-dots as
deterministic sources of photons to investigate interference between
independent quanta. They clearly showed that interference visibility
for comparable intervals was higher for quantum dots with longer
excitation life-times, i.e. wider single-photon pulses.

A different requirement of self-interference is to have the two
slits simultaneously open at any time that a photon can pass. Its
function is self-evident, because the whole concept hinges on the
ability of a wave to propagate from multiple secondary sources and
generate interaction between its components. Thus, it must be
impossible to get interference fringes with a single slit open at a
time, if quantum self-interference is to prevail as a valid
theoretical concept. In practice, this requirement was violated in a
very convincing manner.

It is known that interference fringes do not form, when two slits
are opened alternatively at a very slow rate. However, it was shown
above that photons do not produce fringes at very low rates of
emission even with both slits open. The only relevant settings are
those, in which quanta from both paths have sufficient opportunities
to interact. Such a set-up was prepared successfully by Sillitto and
Wykes \cite{wykes}, who used an electric shutter to switch on and
off the paths of a Young interferometer. Moreover, they were able to
switch both openings several times before any single photon could
reach the detector, even though the two paths were never open
simultaneously. The estimated rate of emission was low enough to
have no more than one photon at a time in the apparatus ($10^{6}$
quanta/sec for a transit time of $10^{-8}$ sec). However, the paths
were controlled with a birefringent device, which means that
components from the same single pulse could have accessed both
trails. Such components could not have entered at the same time, and
they had to be temporally distinguishable inside the interferometer.
Still, they were close enough to interfere with each other. This, in
our opinion, explains the high quality of the results that were
obtained.

The experiment of Sillitto and Wykes was designed to test the role
of uncertainty. The rate of switching exceeded the speed of the
detector, making it impossible to link a click with a path. This, in
terms of the Copenhagen interpretation, proves that uncertainty in
our knowledge overrides physical properties. Even though photons
must have been distinguishable during propagation through the
interferometer, they were not observable as such. However,
distinguishability should be a matter of principle, independent from
the technological endowment of the observer. It should work
regardless of the presence of the observer, or of its choice of
detector. Moreover, the experiment also showed a strong dependency
of fringe visibility on phase coherence, which was controlled by
adjusting the length of one path. Accordingly, the photon had to
``know'' the technological limits of the observer as well as the
exact length of both paths. Note that the photons also had to
interfere with themselves, without being physically able to access
both paths! In our opinion, the hypothesis of second-order quantum
interaction via transient space-time fluctuations is much more
credible for this experiment. Moreover, there are other tests which
confirm our assessment of self-interference.

A very instructive version of the double-slit experiment was
performed by Basano and Ottonello \cite{basano}. They used two
independent lasers, well isolated from each other, in order to
exclude any interaction between them, or between the photons, prior
to interference. The beams of each source were prepared such as to
access only one of two slits. Thus, every single photon had to pass
through one slit at the most (or be extinguished at the screen).
High visibility interference was achieved. Again, the experiment
allows for speculations that our lack of knowledge somehow ``washed
out'' the physical parameters of individual photons. Other
experiments, nevertheless, are adding up to close this loophole.
Santori \textit{et al.} \cite{santori} have demonstrated photon
bunching with independent deterministic sources. In their set-up,
single photons were emitted on demand and their paths through the
interferometer were well-known. The photons had to have clear
physical properties, even though their identity was lost at the
detectors.

A possible objection to these last examples is that they refer to
experiments with multiple sources. Yet, there is another experiment,
in which Fonseca \textit{et al.} demonstrated the so-called
non-local double-slit interference \cite{fonseca}. Entangled
co-propagating photons from a single source were shown to produce
fringes at independent detectors, without ever crossing paths. As a
result, path knowledge was complete for all the photons that
produced fringes in the coincidence count regime. Again, objections
could be raised that entanglement is a special case that does not
apply to this context. However, it is precisely our claim that all
types of Young interference are multiple-photon phenomena expressing
the same underlying physical process. Entangled quanta must also
behave as localized entities, interacting through their associated
waves. If our conclusion is wrong, then the experiment of Fonseca
\textit{et al.} should work with counter-propagating photons as well
(see below). We hope that such an experiment will be performed in
the near future. In any event, the quoted evidence cannot be
reconciled with the assumption of self-interference. This is a
strong argument in favor of \textit{p}-adic valued probability
analysis of quantum interference, which predicted these kinds of
findings. Theory and experiment appear to converge on the conclusion
that double-slit distributions are not reducible to single-entity
phenomena.

\section{Discussion}

The wave properties of optical rays can be explained with great
accuracy by invoking Huygens' Principle. According to the latter,
every point on a wave-front can be treated as a source of secondary
waves. On the downside, this approach was known to have difficulties
explaining the well-defined directions of light waves, as well as
the quantized nature of electromagnetic energy. However, if quanta
are treated as localized propagating particles, generating
space-time perturbations, then both of these problems can be
circumvented. Particles become the unobservable causal substratum
for the electromagnetic waves. (The implication is that
electromagnetic interactions could be analyzed in terms of dynamic
effects on the curvature of space-time). This could also explain the
correlations between detection events in a double-slit experiment,
as suggested by Khrennikov's interpretation of negative
probabilities. Quanta could generate discrete detections, while
still being under the influence of each other's waves, without
violating classical causality. This hypothesis is quite easy to
verify.

Pure waves can generate fringes only within their interference
volume. Guided quanta, on the other hand, must maintain their
momentum even beyond such regions. Is there any reason to believe
that fringes are observable beyond interference volumes? In our
assessment, this phenomenon is fully demonstrated by the so-called
position correlations, displayed by pairs of co-propagating photons
during their coincident detection at separate detectors. As
demonstrated for the first time by Hanbury-Brown and Twiss (HBT),
such correlations are observable at arbitrary distances from the
area of co-propagation, and show up even at unequal distances
between the detectors and the beam-splitter \cite{hbt}. The
experiment of HBT is widely believed to be a consequence of
classical intensity correlations. However, this hypothesis has very
specific experimental consequences, which did not materialize in a
recent attempt to verify them \cite{scarc1}.

Position correlations are an essential element of quantum imaging.
The latter phenomenon was also believed to depend on entanglement at
the source. However, several recent experiments have demonstrated
quantum imaging with chaotic sources of light
\cite{benn1,benn2,scarc2,scarc3}. It should be noted that
experiments with chaotic sources of light were only able to produce
quantum effects with low visibility. Still, they raise an important
question. Is entanglement essential for quantum effects, or is it
just a superior technique for generating reproducible states? We
would like to propose an experimental solution to this problem.
According to our interpretation, fringe build-up requires (roughly)
co-propagating photons. To the best of our knowledge, quantum
imaging, quantum erasure, and other interference-dependent phenomena
have only been demonstrated with set-ups that began with
co-propagating photons. Accordingly, counter-propagating entangled
photons should not be able to reproduce these effects with the same
visibility. At the same time, even independent chaotic beams should
be able to produce quantum effects in co-propagating arrangements.
Moreover, we anticipate the possibility of achieving high-visibility
quantum imaging with chaotic sources of light, well above the
current 30 percent ceiling. These proposals can be fulfilled with
technology that is already available in many labs. Given the high
interest in this topic, we hope that such experiments will be
attempted in the nearest future.

The theoretical and experimental developments discussed in this
paper are opening a new perspective on the nature of quantum
interference. On the one hand, we have better means to interpret the
peculiarities of quantum statistics. On the other hand, we have the
opportunity to test these implications with unprecedented accuracy.
It is still too soon to claim that a final answer on this issue is
available. However, we are persuaded that negative probabilities do
not reflect any kind of anomaly in Nature. On the contrary, it is
our understanding of Nature that has to improve until the mystery is
solved.

\begin{theacknowledgments}
The author is very grateful to Serafino Cerulli-Irelli, Carlos
Monken, Shahriar Afshar, Paul Kwiat, Greg Jaeger, Steve Walborn,
Yoon-ho Kim and Giuliano Scarcelli for their help in finding
relevant data.
\end{theacknowledgments}

\bibliographystyle{aipproc}

\end{document}